\documentclass[prl,aps,twocolumn,showpacs,preprintnumbers,amsmath,amssymb]{revtex4}
\usepackage[dvips]{graphicx}
\usepackage[]{caption}
\usepackage{amsmath}
\usepackage{amssymb}

\usepackage{graphicx}
\usepackage{dcolumn}
\usepackage{bm}
\def\be{\begin{equation}}
\def\ee{\end{equation}}
\def\bea{\begin{eqnarray}}
\def\eea{\end{eqnarray}}

\begin{document}

\preprint{HUTP-06/A0012}

\date{\today}

\title{Tensor Modes from a Primordial Hagedorn Phase of
String Cosmology}

\author{Robert H. Brandenberger$^{1)}$} \email[email: ]{rhb@hep.physics.mcgill.ca}
\author{Ali Nayeri$^{2}$} \email[email: ]{nayeri@schwinger.harvard.edu}
\author{Subodh P. Patil $^{1)}$} \email[email: ]{patil@hep.physics.mcgill.ca}
\author{Cumrun Vafa $^{2)}$} \email[email: ]{vafa@physics.harvard.edu}

\affiliation{1) Dept.of Physics, McGill University, Montr\'eal QC, H3A 2T8, Canada}
\affiliation{2) Jefferson Physical Laboratory, Harvard University, Cambridge,
MA 02138, U.S.A.}


\begin{abstract}
It has recently been shown that a Hagedorn phase of string gas
cosmology can provide a causal mechanism for generating a nearly
scale-invariant spectrum of scalar metric fluctuations, without the
need for an intervening period of de Sitter expansion. In this paper
we compute the spectrum of tensor metric fluctuations (gravitational
waves) in this scenario, and show that it is also nearly
scale-invariant. However, whereas the spectrum of scalar modes has a
small red-tilt, the spectrum of tensor modes has a small blue tilt,
unlike what occurs in slow-roll inflation.  This provides a possible
observational way to distinguish between our cosmological scenario
and conventional slow-roll inflation.
\end{abstract}

\maketitle

\newcommand{\eq}[2]{\begin{equation}\label{#1}{#2}\end{equation}}


{\bf Introduction.} String gas cosmology (SGC) \cite{BV,TV} (see
also \cite{Perlt} for early work, \cite{rbr1,rbr2} for recent
overviews, and \cite{str} for a critical review) is an early
approach to string cosmology, based on adding minimal but crucial
inputs from string theory, namely new degrees of freedom - string
oscillatory and winding modes - and new symmetries - T-duality - to
the now standard hypothesis of a hot and small early universe. One
key aspect of SGC is that the temperature cannot exceed a limiting
temperature, the Hagedorn temperature $T_H$ \cite{Hagedorn}. This
immediately provides a qualitative reason which leads us to expect
that string theory can resolve cosmological singularities \cite{BV}.
From the equations of motion of string gas cosmology \cite{TV,Ven}
it in fact follows that if we follow the radiation-dominated
Friedmann-Robertson-Walker (FRW) phase of standard cosmology into
the past, a smooth transition to a quasi-static Hagedorn phase will
occur. In this phase, the string metric is quasi-static while the
dilaton is time-dependent. Reversing the time direction in this
argument, we can set up the following new cosmological scenario
\cite{BV}: the universe starts out in a quasi-static Hagedorn phase
during which thermal equilibrium can be established over a large
scale (a scale sufficiently large for our current universe to grow
out of it following the usual non-inflationary cosmological
dynamics). The quasi-static phase, however, is not a stable fixed
point of the dynamics, and eventually a smooth transition to a
radiation-dominated FRW phase will occur, after which point the
universe evolves as in standard cosmology.

Recently \cite{Ali1,Ali2} it was discovered that string thermodynamic fluctuations
during the Hagedorn phase lead to scalar metric fluctuations which are
adiabatic and nearly scale-invariant at late times, thus providing a simple alternative
to slow-roll inflation for establishing such perturbations. It is
to be emphasized that this mechanism for the generation of the primordial
perturbations is intrinsically stringy - particle thermodynamic fluctuations
would lead to a spectrum with a large and phenomenologically unacceptable
blue tilt.

We briefly remind the reader of the key features of the new structure
formation scenario. The Figure shows a sketch of the space-time evolution:
at early times $t$ ($ t < t_R$) the universe is in the quasi-static Hagedorn
phase. The physical wavelength of any perturbation mode (characterized
by having constant momentum $k$ in comoving coordinates) is approximately constant.
The key point is that the Hubble radius (which sets the limit on which causal
processes can locally set up fluctuations - see e.g. \cite{RHBrev} for
a concise overview of the theory of cosmological perturbations and
\cite{MFB} for a comprehensive review) is essentially infinite, thus allowing
a causal mechanism for the generation of primordial fluctuations.
Near $t= t_R$, a smooth transition from the Hagedorn phase to the radiation-dominated
phase of standard cosmology occurs. The Hubble radius decreases rapidly to take
on a minimal value which is microscopic (set by the temperature at $t = t_R$
which will be close to the Hagedorn temperature). Thus, fluctuation modes
of relevance to current cosmological observations exit the Hubble radius at
times $t_i(k)$ close to $t_R$. During the radiation-dominated FRW phase,
the Hubble radius increases linearly in $t$, while the physical wavelength
of a perturbation mode grows only as $t^{1/2}$. Thus, at late times $t_{f}(k)$
the fluctuation modes will re-enter the Hubble radius. Since the
primordial perturbations in our scenario are of thermal origin (and
there are no non-vanishing chemical potentials), they will be
adiabatic, and since they propagate on super-Hubble scales for a long time
during the FRW phase, they will be squeezed and will lead to the same
type of acoustic oscillations in the angular power spectrum of the cosmic
microwave background (CMB) anisotropies as what is produced in slow-roll
inflation models.

In this Letter, we generalize our previous analysis \cite{Ali1,Ali2}
to allow us to compute, in addition to the scalar metric
fluctuations (the metric perturbation modes which transform as
scalars under spatial rotations and which couple to matter), also
the tensor modes (gravitational waves). We find that the resulting
spectrum of tensor modes is also nearly scale-invariant, but that it
has a slightly blue tilt, unlike what happens in slow-roll inflation
where the tilt for the gravitational wave spectrum is also red. The
scalar to tensor ratio is calculable from the detailed dynamics of
the system (we postpone this calculation to a followup paper). It is
set by how close the temperature is to the Hagedorn temperature when
the scales which are measured today exit the Hubble radius at the
end of the quasi-static Hagedorn phase. It is not suppressed by the
same small parameter which sets the amplitude of the spectrum of
scalar metric fluctuations. Thus, a measurement of the background of
stochastic gravitational waves would allow us to distinguish our
scenario from the usual slow-roll inflationary models.

The outline of this Letter is as follows: we begin with a
quick recap of the method of \cite{Ali1,Ali2} to compute mass
fluctuations in the Hagedorn phase of SGC and generalize the
method to yield fluctuations of arbitrary components of the
energy-momentum tensor of the string gas. Then, we relate the
spectrum of gravitational waves in late time cosmology to the
fluctuations of the energy-momentum tensor in the Hagedorn phase and show that
the resulting power spectrum is nearly scale-invariant, like the
spectrum of the scalar metric fluctuations.

In the following, we assume that our three spatial dimensions are
already large during the Hagedorn phase (for a possible mechanism to
achieve this see \cite{Natalia1}), while the extra spatial
dimensions are confined to the string scale. For a mechanism to
achieve this in the context of SGC see \cite{BV} (see however
\cite{Col,Frey} for caveats), and for a natural dynamical mechanism
arising from SGC to stabilize all of the moduli associated with the
extra spatial dimensions see
\cite{Watson,Patil1,Patil2,Patil3,Edna,shiggs}. To be specific, we
take our three dimensions to be toroidal. The existence of one
cycles results in the stability of string winding modes - and this
is a key ingredient in our calculations.


{\bf Energy-Momentum Tensor Correlation Functions for Closed
Strings.} We begin with the thermal canonical partition function $Z$
of a gas of closed strings, from which the mean energy-momentum
tensor $\langle T^\mu{}_\nu \rangle$ can be found by 
\be
\langle T^\mu{}_\nu \rangle \, = \, 2 \frac{G^{\mu \alpha}}{\sqrt{ -
G}}\frac{\partial \ln{Z}}{\partial G^{\alpha\nu}}\,
\equiv \, {\cal D}^{\mu}_{\nu} \ln{Z} \, ,
\ee
where $G_{\mu\nu}$ is the Euclidean metric of space-time (the time coordinate
is compactified to a circle of radius $\beta$, where $\beta$ is the inverse
temperature.

Our aim is to calculate the fluctuations of the energy-momentum tensor
on various length scales $R$. For each value of $R$, we will take our
spatial coordinates to run over a fixed interval, e.g. $[0, 2\pi]$. Thus,
the metric will have dimensions and will be given by
\be \label{metric}
G_{\mu \nu} \, = \, {\rm diag}[\beta^2, R^2, R^2, R^2] \, .
\ee
Letting another derivative operator ${\cal D}$ act on $\ln{Z}$ gives
two terms, one for which both derivative operators act on $Z$, a
second which will contain the product of terms where one ${\cal D}$
acts on $Z$. Specifically (and symmetrizing over the indices), we
find the mean square fluctuation to be
\bea
C^\mu{}_\nu{}^\sigma{}_\lambda & = & \langle \delta T^\mu{}_\nu
\delta T^\sigma{}_\lambda \rangle  =  \langle T^\mu{}_\nu
T^\sigma{}_\lambda \rangle  - \langle T^\mu{}_\nu \rangle \langle
T^\sigma{}_\lambda \rangle  \nonumber \\
& = & 2 \frac{G^{\mu \alpha}}{\sqrt{ - G}}\frac{\partial}{\partial
G^{\alpha\nu}}\left(\frac{G^{\sigma \delta}}{\sqrt{ -
G}}\frac{\partial \ln{Z}}{\partial G^{\delta\lambda}}\right) \nonumber \\
&+& 2 \frac{G^{\sigma \alpha}}{\sqrt{ - G}}\frac{\partial}{\partial
G^{\alpha\lambda}}\left(\frac{G^{\mu \delta}}{\sqrt{ -
G}}\frac{\partial \ln{Z}}{\partial G^{\delta\nu}}\right) \,,
\eea
with $\delta T^\mu{}_\nu = T^\mu{}_\nu - \langle T^\mu{}_\nu
\rangle$.

The partition function $Z = \exp{(-\beta F)}$ is given by the string
free energy $F$. Thus, the string thermodynamical fluctuation in the
energy density, here denoted by the correlation function
$C^0{}_0{}^0{}_0$, becomes
\bea
C^0{}_0{}^0{}_0 \, &=& \, \langle \delta\rho^2 \rangle = \,
\langle \rho^2
\rangle - \langle \rho \rangle ^2 \nonumber \\
&=& \, - \frac{1}{R^{6}} \frac{\partial}{\partial \beta}\left(F +
\beta \frac{\partial F}{\partial \beta}\right) \, = \,
\frac{T^2}{R^6} C_V \, . \label{cor1}
\eea
where $C_V = (\partial \langle E \rangle / \partial T)$  with $E
\equiv F + \beta (\partial F/\partial \beta)$ and $V = R^3$ is the
volume of three compact but large spatial dimensions. The
fluctuation in spatial diagonal parts of the energy-momentum tensor
can be found by 
\bea \label{cor2} C^i{}_i{}^i{}_i & = & \langle \delta {T^i{}_i}^2
\rangle = \langle {T^i{}_i}^2 \rangle - \langle T^i{}_i \rangle^2
\nonumber \\
& = & \frac{1}{\beta R^3}\frac{\partial}{\partial \ln{R}}\left(-
\frac{1}{R^3} \frac{\partial F}{\partial \ln{R}}\right)  =
\frac{1}{\beta R^2}\frac{\partial p}{\partial R} \,, \eea
with no summation on $i$. The pressure $p$ is given by
\be
p \,  \equiv   \, - V^{-1}(\partial F/\partial \ln{R}) \, = \, T
(\partial S/\partial V)_E  \, . \label{stringypressure}
\ee

In the following, we will compute the two correlation functions
(\ref{cor1}) and (\ref{cor2}) using tools from string statistical
mechanics. Specifically, we will be following the discussion in
\cite{Deo} (see also \cite{Deo1,Deo2,mb,mt}). The starting point is
the formula $S(E , R) \, = \, \ln{\Omega(E ,R)}$ for the entropy in
terms of $\Omega(E ,R)$, the density of states. The density of
states of a gas of closed strings on a large three-dimensional torus
(with the radii of all internal dimensions at the string scale) was
calculated in \cite{Deo} and is given by 
\be
\Omega(E , R) \, \simeq \, \beta_H e^{\beta_H E + n_H V}[1 +
\delta \Omega_{(1)}(E , R)] \label{density_states}\,,
\ee
where $\delta \Omega_{(1)}$ comes from the contribution to the density of states
(when writing the density of states as a Laplace transform of $Z(\beta)$,
which involves integration over $\beta$) from the closest singularity point
$\beta_1$ to $\beta_H = (1/T_H)$ in the complex $\beta$ plane. Note that
$\beta_1 < \beta_H$, and $\beta_1$ is real. From \cite{Deo} we have
\be
\delta \Omega_{(1)}(E , R) \, = \, - \frac{(\beta_H E)^{5}}{5!}
e^{-(\beta_H - \beta_1)(E  - \rho_H V)} \,.
\ee
In the above, $n_H$ is a (constant) number density of order
$\ell_s^{-3}$ ($\ell_s$ being the string length) and $\rho_H$ is the
`Hagedorn Energy density' of the order $\ell_s^{-4}$, and
\be
\beta_H - \beta_1 \sim \left\{ \begin{array}{ll}
(\ell_s^3/R^2) \,, & \mbox{for $R \gg \ell_s$}\,, \\
(R^2/\ell_s)\,, & \mbox{for $R \ll \ell_s$}\,.
\end{array}
\right.
\ee
To ensure the validity of Eq. (\ref{density_states}) we demand that
$\delta \Omega_{(1)} \ll 1$ by assuming $\rho \equiv (E / V) \gg
\rho_H$.

Combining the above results, we find that he entropy of the string gas
in the Hagedorn phase is given by
\be
\label{entropy} S(E , R) \simeq \beta_H E + n_H V + \ln{\left[1
+ \delta \Omega_{(1)}\right]} \,,
\ee
and therefore the temperature $T (E, R) \equiv [(\partial S/\partial
E)_V]^{-1}$ will be
\be 
T  \simeq  \left(\beta_H + \frac{\partial \delta
\Omega_{(1)}/\partial E}{1 + \delta \Omega_{(1)}}\right)^{-1} \simeq
T_H \left(1 + \frac{\beta_H - \beta_1}{\beta_H} \delta
\Omega_{(1)}\right)\label{temp}\,. 
\ee 
Using this relation, we can express $\delta \Omega_{(1)}$ in terms of
$T$ and $R$ via
\be
\label{deltaomega}
\ell_s^3\delta \Omega_{(1)} \, \simeq \, - \frac{R^2}{T_H}\left(1 -
\frac{T}{T_H}\right) \, .
\ee
In addition, we find
\be \label{meanen}
\langle E \rangle \, \simeq \,  \ell_s^{-3} R^2
\ln{\left[\frac{\ell_s^3 T}{R^2 (1- T/T_H)}\right]}\,.
\ee
Note that to ensure that $\delta \Omega_{(1)} \ll 1$ and $\langle E
\rangle  \gg \rho_H R^3$, one should demand $(1 - T/T_H) \gg
(\ell_s^2/R^2)$.

The results (\ref{entropy}) and (\ref{deltaomega}) now allow us
to compute the correlation functions (\ref{cor1}) and (\ref{cor2}).
We first compute the energy correlation function (\ref{cor1}),
noting that
\be
C_V \equiv -\left[T^2 \left(\frac{\partial^2 S(E , R)}{\partial
E^2}\right)_V\right]^{-1} \approx \frac{R^2/\ell^3}{T \left(1 -
T/T_H\right)}\,,
\ee
from which we get
\be
C^0{}_0{}^0{}_0 = \langle\delta \rho^2\rangle\, \simeq \,
\frac{T}{\ell_s^3(1 - T/T_H)} \frac{1}{R^4}\, .
\ee
Note that the factor $(1 - T/T_H)$ in the denominator is responsible
for giving the spectrum a slight red tilt. It comes from the differentiation
with respect to $T$.

Evaluating (\ref{stringypressure})
\be
p(E, R) \approx n_H T_H - \frac{2}{3}\frac{(1 - T/T_H)}{\ell_s^3
R}\ln{\left[\frac{\ell_s^3 T}{R^2 (1- T/T_H)}\right]} \,,
\ee
immediately yields
\be C^i{}_i{}^i{}_i \, \simeq \, \frac{T (1 - T/T_H)}{\ell_s^3 R^4}
\ln^2{\left[\frac{R^2}{\ell_s^2}(1 - T/T_H)\right]}\, . \ee
Note that since no temperature derivative is taken, the factor $(1 - T/T_H)$
remains in the numerator, and it is this fact which will lead to the
slight blue tilt of the spectrum of gravitational waves.


{\bf Tensor Modes from Hagedorn Fluctuations.} In this section we
estimate the dimensionless power spectrum of gravitational waves.
First, we make some general comments. In slow-roll inflation, to
leading order in perturbation theory matter fluctuations do not
couple of tensor modes. This is due to the fact that the matter
background field is slowly evolving in time and the leading order
gravitational fluctuations are linear in the matter fluctuations. In
our case, the background is not evolving (at least at the level of
our computations), and hence the dominant metric fluctuations are
quadratic in the matter field fluctuations. At this level, matter
fluctuations induce both scalar and tensor metric fluctuations.
Based on this consideration we expect that in our string gas
cosmology scenario, the ratio of tensor to scalar metric
fluctuations will be larger than in simple slow-roll inflationary
models.

We will extract the amplitude of the gravitational wave spectrum
from the spatial fluctuations $C^i{}_j{}^i{}_j$
of the energy-momentum tensor. Strictly speaking, it is the
off-diagonal components which will couple uniquely to the tensor
modes. We will estimate their order of magnitude by the order of
magnitude of the diagonal terms computed in the previous section.
This gives a good approximation, as can be checked
by letting the metric in (\ref{metric})
depend on three separate scales $R_i$ (where the index $i$ runs from
1 to 3), and by extracting the off-diagonal correlation functions
following the method of the previous section, but taking mixed spatial
derivatives.

Tensor perturbations in a spatially flat FRW universe take the form
\be
d s^2 = - d t^2 + a^2(t)(\delta_{ij} + h_{ij})d x^i d x^j \,.
\ee
Since to linear order in $h_{ij}$ the Einstein tensor for fluctuations
on a scale $k$ is proportional to $k^2 h_{ij}(k)$, it follows from
the space-space Einstein equations that
\be \label{pertEE}
k^2 h_{ij}(k) \, \sim \, (8 \pi G) \delta T_{ij}(k) \, .
\ee
The power spectrum of the right-hand side of the above equation is given
by the correlation function $C^i{}_j{}^i{}_j$. Thus, the dimensionless
gravitational wave power spectrum is given by this correlation function.
Therefore, from (\ref{pertEE}) one can calculate the dimensionless
power spectrum
for $h^{\pm}_k$, where $h^{\pm}_k$ is the amplitude of either of the
two gravitational wave polarization modes. Dropping the superscript
$\pm$ (due to symmetry both polarization modes will be equally excited)
we obtain
\be
k^3 |h(k)|^2 \, \sim \, k^{-4} (8 \pi G)^2 C^i{}_j{}^i{}_j \, .
\ee
Inserting our result (\ref{cor2}) for the correlation function
yields
\be
k^3 |h(k)|^2 \, \sim \, {{64 \pi^2 G^2 T} \over {\ell_s^3}} (1 -
T/T_H) \ln^2{\left[\frac{1}{\ell_s^2 k^2}(1 - T/T_H)\right]}\, , \ee
which, for temperatures close to the Hagedorn value reduces to
\be \label{tensorresult} k^3 |h(k)|^2 \, \sim \,
\left(\frac{\ell_{Pl}}{\ell_s}\right)^4 (1 -
T/T_H)\ln^2{\left[\frac{1}{\ell_s^2 k^2}(1 - T/T_H)\right]} \, . \ee
This shows that the spectrum of tensor modes is - to a first approximation -
scale-invariant.


{\bf Discussion.}  Our result (\ref{tensorresult}) for the power
spectrum of gravitational waves should be compared to the result for
the power spectrum of scalar metric fluctuations computed in
\cite{Ali1}: 
\be \label{scalarresult} {\cal{P}}_\Phi(k) \, \sim \,
\left({{\ell_{pl}} \over {\ell_s}}\right)^4 {1 \over {1 - T/T_H}} \,
. \ee
Note that for a fixed scale $k$, both (\ref{scalarresult}) and
(\ref{tensorresult}) must be evaluated at the time $t_i(k)$ when the
mode $k$ exits the Hubble radius at the end of the Hagedorn phase.
Since $t_i(k)$ increases slightly as $k$ increases, the temperature
$T(t_i(k))$ will be slowly decreasing. Hence, the expression in front
of the logarithm in our final expression (\ref{tensorresult}) for the
power spectrum of tensor fluctuations yields a slight blue tilt.
Values of $k$ for which the perturbative analysis of string gas
cosmology is consistent are on the high $k$ side of the zero of the
logarithm in (\ref{tensorresult}) - this follows from the
condition $ |\delta \Omega_{(1)}| \ll 1$ and (\ref{deltaomega}). Hence,
the logarithm factor in (\ref{tensorresult}) adds to the blue tilt
of the spectrum.  

A heuristic way of understanding the origin of the slight blue tilt
in the spectrum of tensor modes 
is as follows. The closer we get to the Hagedorn temperature, the
more the thermal bath is dominated by long string states, and thus
the smaller the pressure will be compared to the pressure of a pure
radiation bath. Since the pressure terms (strictly speaking the
anisotropic pressure terms) in the energy-momentum tensor are
responsible for the tensor modes, we conclude that the smaller the
value of the wavenumber $k$ (and thus the higher the temperature
$T(t_i(k))$ when the mode exits the Hubble radius, the lower the
amplitude of the tensor modes. In contrast, the scalar modes are
determined by the energy density, which increases at $T(t_i(k))$ as
$k$ decreases, leading to a slight red tilt.

It is also interesting to consider the tensor to scalar ratio $r$.
Comparing (\ref{tensorresult}) and (\ref{scalarresult}) we see that
this ratio, evaluated on a scale $k$ is given by
\be
r \, \sim \, (1 - T(t_i(k))/T_H)^2
\ln^2{\left[\frac{1}{\ell_s^2 k^2}(1 - T(t_i(k))/T_H)\right]}\, .
\ee
In principle (if the dynamical evolution from the Hagedorn phase to
the radiation-dominated FRW phase were under complete analytical
control) this quantity is calculable. If the string length were
known, the factor $(1 - T/T_H)$ could be determined from the
normalization of the power spectrum of scalar metric fluctuations.
Since the string length is expected to be a couple of orders larger
than the Planck length, the above factor does not need to be very
small. Thus, generically we seem to predict a ratio $r$ larger than
in simple roll inflationary models.

Based on the results of this Letter it thus appears promising that
our scenario will give rise to testable predictions.

\begin{acknowledgments}

The work of R.B. is supported by funds from McGill University, by an
NSERC Discovery Grant and by the Canada Research Chairs program. The
work of A.N. and C.V. is supported in part by NSF grant PHY-0244821
and DMS-0244464.

\end{acknowledgments}

\end{document}